\documentclass[prd,twocolumn,showpacs,nofootinbib]{revtex4}
\usepackage{times}
\usepackage{natbib}
\usepackage{epsfig}
\usepackage{bm}
\usepackage{comment}
\usepackage{color}

\newcommand{\bdv}[1]{{\bf #1}}
\newcommand{\np}{n_g}
\newcommand{\msun}{{\it h}^{-1}M_{\odot}}

\newcommand{\beeq}{\begin{equation}}
\newcommand{\bnobs}{\bar\np}
\newcommand{\zz}{z}      
    
\newcommand{\dz}{\delta z}           
\newcommand{\dzv}{\delta z_v}
\newcommand{\dv}{\delta_v}
\newcommand{\ee}{e}                 
\newcommand{\bee}{\boldsymbol{\ee}}
\newcommand{\bpp}{\boldsymbol{p}}
\newcommand{\eneq}{\end{equation}}
\newcommand{\dgr}{\delta_\up{obs}}
\newcommand{\ax}{\alpha_{\chi}}     
\newcommand{\px}{\varphi_{\chi}}    
\newcommand{\dzg}{\dz_\chi}
\newcommand{\drg}{\delta r_{\chi}}  
\newcommand{\rs}{r}
\newcommand{\HH}{\mathcal{H}}
\newcommand{\ddL}{\delta\mathcal{D}_L} 
\newcommand{\kag}{\mathcal{K}}      
\newcommand{\bear}{\begin{eqnarray}}
\newcommand{\vx}{v_{\chi}}          

\newcommand{\dnewt}{\delta_\up{Newt}}
\newcommand{\OM}{\Omega_m}
\newcommand{\cA}{\mathcal{P}}    
\newcommand{\cB}{\mathcal{R}}    
\newcommand{\enar}{\end{eqnarray}}
\newcommand{\bdobs}{\boldsymbol{\delta}_\up{GR}^\up{obs}}
\newcommand{\bdobsT}{\boldsymbol{\delta}_\up{GR}^\up{obs\dagger}}
\newcommand{\bb}{\boldsymbol{b}_0}         
\newcommand{\bI}{\boldsymbol{I}}    
\newcommand{\bnn}{\boldsymbol{\varepsilon}}   
\newcommand{\BB}{\boldsymbol{b}}       
\newcommand{\up}[1]{{\rm #1}}
\newcommand{\CCI}{\bdv{C}^{-1}} 
\newcommand{\CC}{\bdv{C}} 
\newcommand{\AVE}[1]{\langle#1\rangle}
\newcommand{\BT}{\boldsymbol{b}^\dagger}     
\newcommand{\Pv}{P_m}
\newcommand{\EE}{\boldsymbol{\mathcal{E}}}       
\newcommand{\mA}{\alpha}   
\newcommand{\EEI}{\boldsymbol{\mathcal{E}}^{-1}} 
\newcommand{\mB}{\beta}   
\newcommand{\mBS}{\beta^*}   

\newcommand{\mC}{\gamma}   
\newcommand{\fnl}{f_\up{NL}}
\newcommand{\gpc}{\rm Gpc}
\newcommand{\hgpc}{{h^{-1}\gpc}}
\newcommand{\hmpci}{h\up{Mpc}^{-1}}

\newcommand{\hmsun}{h^{-1}M_\odot}
\newcommand{\DD}{{m_{\dz}}}

\newcommand{\kmin}{k_\up{min}}
\newcommand{\kmax}{k_\up{max}}
\newcommand{\Mmin}{M_\up{min}}

\newcommand{\ccA}{c_{\cA}}
\newcommand{\ccB}{c_{\cB}}

\newcommand{\cp}{\varphi_v}
\newcommand{\dm}{\delta_m}            
\newcommand{\VV}{\mathcal{V}}

\newcommand{\kvec}{\bdv{k}}
\newcommand{\xvec}{\bdv{x}}

\begin{document}

\title{Going beyond the Kaiser redshift-space distortion formula:
a full general relativistic account of the 
effects and their detectability in galaxy clustering}

\author{Jaiyul Yoo$^{1,2}$}
\altaffiliation{jyoo@physik.uzh.ch}
\author{Nico Hamaus$^1$}
\author{Uro{\v s} Seljak$^{1,2,3,4}$}
\author{Matias Zaldarriaga$^{5}$}
\affiliation{$^1$Institute for Theoretical Physics, University of Z\"urich,
CH-8057 Z\"urich, Switzerland}
\affiliation{$^2$Lawrence Berkeley National Laboratory, University of 
California, Berkeley, CA 94720, USA}
\affiliation{$^3$Physics Department and Astronomy Department,
University of California, Berkeley, CA 94720, USA}
\affiliation{$^4$Institute for the Early Universe, Ewha Womans University, 
120-750 Seoul, South Korea}
\affiliation{$^5$School of Natural Sciences, Institute for Advanced Study, 
Einstein Drive, Princeton, NJ 08540, USA}

\begin{abstract}
Kaiser redshift-space distortion formula describes well the clustering of 
galaxies in 
redshift surveys on small scales, but there are numerous additional terms
 that arise 
on large scales. Some of these terms can be described using Newtonian 
dynamics
and have been discussed in the literature, while the
others require proper general relativistic description that was only
 recently developed.  
Accounting for these terms in galaxy clustering is the first step toward 
tests of general relativity on horizon scales.
The effects can be 
classified as two terms that represent the velocity and the gravitational
potential contributions. Their amplitude is determined by 
effects such as the volume and luminosity distance fluctuation effects and the 
time evolution of galaxy number density and 
Hubble parameter. We compare the Newtonian approximation 
often used in the redshift-space distortion literature
to the fully general relativistic equation, and show that Newtonian 
approximation 
accounts for most of the terms contributing to velocity effect.
We perform a Fisher matrix analysis of detectability of these terms and show 
that in a 
single tracer survey they are completely undetectable. To detect these
 terms one 
must resort to the recently developed methods to reduce sampling variance and 
shot noise. We show that in an all-sky galaxy redshift survey at low
redshift the velocity term can be measured at a few sigma
if one can utilize halos of mass 
$M\geq10^{12}\msun$ (this can increase to 10-$\sigma$ or more in some more
 optimistic scenarios),
while the gravitational potential term itself can only be marginally detected.
We also demonstrate that the general relativistic effect is not
degenerate with the primordial non-Gaussian signature in galaxy bias,
and the ability to detect primordial non-Gaussianity is 
little compromised.
\end{abstract}


\maketitle

\section{Introduction}
In the past few decades galaxy redshift surveys have been one of the 
indispensable
tools in cosmology, covering a progressively larger fraction of the sky
with increasing redshift depth. 
With the upcoming dark energy surveys this trend
will continue in the future. However, 
despite the advance in observational
frontiers, there remained a few unanswered questions in 
the theoretical description
of galaxy clustering. One is the issue of validity of the Kaiser formula, 
where density perturbation in redshift space is density perturbation in 
real space multiplied with a term that depends on the angle between the 
line-of-sight direction and the direction of the Fourier mode (we give a 
more detailed definition below) \cite{KAISE87}. 
It has been well known (e.g., \cite{HAMIL98})
that the simplest version omits some of the terms coming from the
Jacobian of the transformation from real space to redshift space, terms of 
order $v/\HH r$, where $v$ is the velocity and $r$ is the distance 
to the galaxy and $\HH$ is the conformal Hubble parameter.
It is argued that these terms are potentially important, especially for 
large angular separations (e.g., see,
\cite{SZMALA98,SZAPU04,PASZ08,RASAPE10,BEMAET12}),
but most of the analyses so far have focused on effects in correlation 
function, without proper signal-to-noise analysis (see, however, 
\cite{SAPERA12}).

A second, related issue, is whether the terms originally derived in 
the Newtonian 
approximation, get modified when a proper general relativistic description 
is employed. 
On horizon scales, the standard Newtonian
description naturally breaks down, and a choice of hypersurface of simultaneity
becomes an inevitable issue, demanding a fully relativistic treatment
of galaxy clustering beyond the current Newtonian description.
In recent work \cite{YOFIZA09,YOO10}, it is shown that a proper relativistic
description can be easily obtained by following the observational procedure
in constructing the galaxy fluctuation field and its statistics: We need to
model observable quantities, rather than theoretically convenient but 
unobservable quantities, usually adopted in the standard method.
While both the relativistic and the standard Newtonian
descriptions are virtually indistinguishable in the Newtonian
limit, they are substantially different on horizon scales, rendering galaxy
clustering measurements a potential probe of general relativity. 

The relativistic description of galaxy clustering includes 
two new terms that scale as velocity and gravitational potential. 
Compared to the dominant density contribution, they are suppressed by
$\HH/k$ and $(\HH/k)^2$ and become important only on large scales, where
the comoving wavevector amplitude is~$k$. Consequently, the
identification of these terms just by looking at the galaxy power spectrum is
hampered  because 
of sampling variance, and the general relativistic
effects unaccounted in the standard Newtonian description may result in
systematic errors less than 1-$\sigma$ for most 
of the volume available at $z\leq3$ in the standard power spectrum analysis
\cite{YOO10}. We revisit this calculation here, using a more realistic 
description 
of these terms, but the basic conclusion remains the same. 

In light of the fact that these corrections to the Kaiser formula provide a 
potential probe of the consistency of our model, including generic tests 
of general 
relativity on large scales, it is worth exploring if these terms can be 
observed. 
A new multi-tracer method \cite{SELJA09} takes advantage of the fact 
that differently biased galaxies 
trace the same underlying matter distribution, and it can be
used to cancel the randomness of the matter distribution in a single 
realization of the Universe, eliminating the sampling variance limitation. 
This method has been used in \cite{MCDON09} to investigate the velocity 
effects 
of \cite{YOFIZA09,YOO10}, noting that for any given Fourier mode
the imaginary part of velocity couples to the real part of density and vice 
versa. 
Even with this novel technique, the expected detection level is low 
\cite{MCDON09}.

If sampling variance is eliminated, the dominant remaining source of error 
is shot noise, 
caused by the discrete nature of galaxies. Recently, a shot noise cancelling 
technique has been proposed \cite{SEHADE09} and investigated for detecting 
primordial non-Gaussianity \cite{HASEDE11}
in combination with the sampling variance cancelling technique.
The basis of the method is that by using halo mass dependent weights one can 
approximate a halo field 
as the dark matter field and reduce the stochasticity between them. While 
this works best when 
comparing halos to dark matter, some shot noise cancelling can also be 
achieved by comparing halos 
to each other \cite{HASEET10,HASEDE11}.
This opens up a new opportunity to probe horizon scales and
extract cosmological information with higher signal-to-noise ratio.

In this work our primary goal is to explore the detectability of 
these effects in galaxy clustering
on cosmological horizon scales using the galaxy power spectrum
measurements with both single tracer technique and with the multi-tracer 
and shot noise cancelling 
techniques. 
In addition we clarify the relation 
of the galaxy fluctuation field
often used in the redshift-space distortion literature using Newtonian 
approximation 
to the fully general relativistic equation.
Finally, we also investigate the impact of the general relativistic effects
in detecting the primordial non-Gaussian signature in galaxy bias.

The organization of the paper is as follows. In Sec.~\ref{sec:formal},
we present the full general relativistic description of galaxy clustering.
In Sec.~\ref{sec:newt}, we discuss the Newtonian correspondence and its
relation to the redshift-space distortion literature.
The multi-tracer and shot noise cancelling techniques are presented in
Sec.~\ref{sec:multi}. In Sec.~\ref{sec:signi}, we provide the measurement
significance of the general relativistic effects in the galaxy power spectrum.
In Sec.~\ref{sec:fnl}, we extend our formalism to the primordial 
non-Gaussianity. Finally, 
we discuss the implication of our results in Sec.~\ref{sec:discussion}. 

\section{General Relativistic Description of Galaxy Clustering} 
\label{sec:formal}
A full general relativistic description of galaxy
clustering is developed in \cite{YOFIZA09,YOO10} (see also
\cite{BODU11,CHLE11,JESCHI11}). 
Previously, we have adopted the simplest linear bias ansatz,
in which the galaxy number density is just a function
of the matter density $\np=F[\rho_m]$, both at the same 
spacetime.\footnote{On large scales, a more general stochastic relation
between the galaxy number density and the matter density also
reduces to the local form we adopted here [11]. As opposed
to some confusion in literature, this biasing scheme is independent of whether
galaxies are observed.}
However, this ansatz turns out to be rather restrictive, since the time 
evolution 
of the galaxy sample is entirely driven by the evolution of the matter density
$\propto(1+z)^3$. Here we make one modification in the adopted linear 
bias ansatz by relaxing this assumption and providing more freedom,
while keeping the locality. We allow the galaxy number density at the same
matter density to differ depending on its local history, as a local observer
at the galaxy formation site
is affected by its local matter density and the proper time to the linear
order in perturbation, i.e., $\np=F[\rho_m,t_p]$ with $t_p$ being
the proper time measured in the galaxy rest frame. 
Physically, the presence of long wavelength modes affects 
the local dynamics of galaxy formation by changing the local curvature
and thus the expansion rate, and these are modulated by the Laplacian of the
comoving curvature and the proper time
\cite{BASEET11}. Therefore, in addition to the contribution of
the matter density fluctuation
$\DD=\delta-3~\dz$ at the observed redshift~$\zz$ \cite{YOFIZA09,YOO10}, 
the physical number density of galaxies has additional contribution from
the distortion between the observed redshift and the proper time, 
when expressed at the observed redshift:
\bear
\np&=&\bnobs(\zz)\left[1+b~(\delta-3~\dz)-b_t~\dzv\right] \nonumber \\
&=&\bnobs(\zz)\left[1+b~\dv-\ee~\dzv\right]~,
\label{eq:bias}
\enar
where the matter density fluctuation is~$\delta$,
the lapse~$\dz$ in the observed redshift~$\zz$ is defined as 
$1+\zz=(1+\bar z)(1+\dz)$, and the homogeneous redshift parameter is 
related to the cosmic expansion factor as~$1/a=1+\bar z$.
The subscript~$v$ indicates quantities are evaluated in the dark matter 
comoving gauge.\footnote{Gauge-dependence arises, 
only when perturbation quantities are considered. For example, the
physical number density of galaxies is a well-defined scalar field without
gauge ambiguity. However, when we split it into the homogeneous part and
the perturbation part, the correspondence of the physical quantity to
the homogeneous part depends on the coordinate choice, and 
consequently the perturbation part becomes gauge-dependent \cite{BARDE80}.
The comoving gauge is a choice of gauge conditions, in which
 $v=0$ or the off-diagonal component of the energy-momentum tensor is zero.
No simple choice of gauge conditions corresponds to the observer's choice
of coordinates ($\zz$, $\theta$, $\phi$).}

The galaxy bias factor is
$b=\partial\ln\bnobs/\partial\ln\rho_m|_{t_p}$ and
the time evolution of the galaxy number density due to its local history is
$b_t=\partial\ln\bnobs/\partial\ln(1+z)|_{\rho_m}$. Therefore, 
the total time evolution of the mean galaxy number density is 
proportional to the evolution bias \cite{YOO09},
\beeq
\ee={d\ln\bnobs\over d\ln(1+z)}=3~b+b_t~.
\label{eq:eee}
\eneq
For galaxy samples with constant comoving number density, the evolution 
bias factor is $e=3$ due to our use of the physical number density in 
Eq.~(\ref{eq:eee}).
This biasing scheme in Eq.~(\ref{eq:bias})
is consistent with \cite{BODU11,CHLE11,BASEET11,BRCRET11,JESCHI11}, 
and our previous bias ansatz corresponds to $b_t=0$. 

Therefore, with this more physically motivated bias ansatz,
the general relativistic description of the observed galaxy 
fluctuation is \cite{YOFIZA09,YOO10}
\bear
\label{eq:fullGR}
\dgr&=&\left[b~\dv-e~\dzv\right]+\ax+2~\px+V+3~\dzg \nonumber \\
&+&2~{\drg\over\rs}
-H{d\over d\zz}\left({\dzg\over\HH}\right)-5p~\ddL-2~\kag~. 
\enar
Here the luminosity function slope of the source galaxy population 
at the threshold is~$p$. Note that this is the slope of the luminosity 
function, expressed in terms of absolute magnitude $M=\up{constant}-2.5 
\log L$, hence the factor 
$5p$ instead of the factor $2\alpha_L$, where $\alpha_L$ is the slope of the 
luminosity function expressed in terms of
 luminosity $L$ [see Eq.~(\ref{eq:slope})]. In addition, 
the comoving line-of-sight distance to the observed redshift
is~$\rs$, the dimensionless fluctuation in the luminosity distance is $\ddL$,
the temporal and spatial metric perturbations are
$\ax$ and $\px$, the line-of-sight velocity is~$V$, and the gauge-invariant
radial displacement and lensing convergence are $\drg$ and 
$\kag$.\footnote{The displacement of the source galaxy position from the
observed galaxy position is split into the radial and angular components,
and these two parts are expressed in terms of gauge-invariant quantities,
i.e., their forms differ in each gauge choice but 
give the same value regardless of gauge choice.
These quantities are most conveniently expressed in the conformal Newtonian
gauge. In \cite{YOO10} the radial distortion is denoted as
$\delta\mathcal{R}$. Here we use a slightly different notation 
for the radial distortion $\drg$ to avoid confusion with the 
dimensionless coefficient~$\cB$ in Eq.~(\ref{eq:PR}).
}
The subscript~$\chi$ indicates quantities are evaluated in 
the conformal Newtonian gauge (also known as the zero-shear $\chi=0$ 
gauge).\footnote{The covariant derivative of the observer velocity is often
decomposed into expansion, shear, rotation, and acceleration vectors
\cite{KOSA84}. The shear component is proportional to $\chi=a(\beta+\gamma')$
with the metric convention in \cite{YOO10}.
The conformal Newtonian gauge or the zero-shear gauge corresponds to
the frame, in which there is no shear seen by a observer moving orthogonal
to the constant-time hypersurface.}
We have ignored the vector and tensor contributions to $\dgr$ in 
Eq.~(\ref{eq:fullGR}).

We emphasize that compared to \cite{YOFIZA09,YOO10}
it is only the terms in the square bracket that are
affected by the choice of linear bias ansatz, and Eq.~(\ref{eq:fullGR})
is consistent with  \cite{BODU11,CHLE11,BASEET11,BRCRET11,JESCHI11}.
Various other terms in Eq.~(\ref{eq:fullGR}) arise due to the mismatch 
between the observed and the physical quantities.
The radial and angular distortions are represented by $\drg$ and $\kag$, and
the distortion in the observed redshift corresponds to the derivative
term. The conversion of physical volume to comoving volume gives rise to
the distortion $3~\dzg$, and the rest of the potential and velocity terms 
defines the local Lorentz frame, where the source galaxies are defined.
For galaxy samples selected by the observed flux, additional contribution
$5p~\ddL$ arises. Further discussion regarding this contribution
is given later in this section.

Since Eq.~(\ref{eq:fullGR}) is written in terms of gauge-invariant variables,
it can be evaluated with any choice of gauge conditions. However, in
evaluating Eq.~(\ref{eq:fullGR}), it proves convenient to use different
combinations of gauge-invariant variables, rather than to choose one specific
gauge condition and convert all the gauge-invariant variables in
Eq.~(\ref{eq:fullGR}) to quantities in the chosen gauge.

For a presureless medium in a flat universe, the Einstein equations are
(e.g., \cite{BARDE80,KOSA84,HWNO01})
\bear
\label{eq:poisson}
k^2\px&=&{3H_0^2\over2}\OM~{\dv\over a}~, \\
\cp'&=&\HH\alpha_v~,\\
\label{eq:pot}
\ax&=&-\px~,
\enar
and the conservation equations are
\bear
\label{eq:cons}
&&\vx'+\HH\vx=k\ax~,\\
&&\dv'=-3\cp'-k\vx~,
\enar
where the prime is a derivative with respect to the conformal time~$\tau$ and
the equations are in Fourier space. From these equations, it is well-known
that the dark-matter comoving gauge ($v=0$) is coincident with synchronous
gauge [$\ax=\alpha_v+(a\vx)'/ak$, hence $\alpha_v=0$ from Eq.~(\ref{eq:cons})]
and the comoving curvature is conserved ($\cp'=0$) \cite{HWNO99R,WASL09}. 
Furthermore, since these gauge-invariant variables correspond to
the usual Newtonian quantities \cite{HWNO99R,HWNO05},
hereafter we adopt a simple notation
$\dv\equiv\dm$, $\vx\equiv v$, and $\px=-\ax\equiv\phi$ to emphasize their
connection.

To facilitate evaluation of Eq.~(\ref{eq:fullGR}), we express the remaining
gauge-invariant quantities in terms of $\dm$, $v$, and $\phi$ as
\bear
\label{eq:los}
V&=&{\partial\over \partial r}
\int{d^3\kvec\over(2\pi)^3}{-v(\kvec)\over k}~
e^{i\kvec\cdot\xvec}~, \\
\label{eq:dz}
\dzg&=&V+\phi+\int_0^r d\tilde r~2\phi'~, \\
\label{eq:dzv}
\dz_v&=&\dzg+\int{d^3\kvec\over(2\pi)^3}~{\HH v\over k}~
e^{i\kvec\cdot\xvec}~, \\
\label{eq:dr}
\drg&=&-{\dzg\over\HH}-\int_0^rd\tilde r~2\phi~,\\
\label{eq:kappa}
\kag&=&-\int_0^rd\tilde r\left({r-\tilde r\over \tilde r r}\right)
\hat\nabla^2\phi~,\\
\label{eq:dl}
\ddL&=&\phi+V-{\dzg\over\HH r}+\int_0^rd\tilde r~
{\tilde r\over r}~2\phi'-{1\over r}\int_0^rd\tilde r~2\phi \nonumber \\
&&+\int_0^rd\tilde r{(r-\tilde r)~\tilde r\over r}
\left(\Delta\phi+\phi''-2{\partial\phi'\over\partial \tilde r}\right)~ 
\nonumber \\
&=&{\drg\over r}+\dzg+\px-\kag~,  \\
\label{eq:zdist}
&&\hspace{-40pt}
-H{d\over dz}\left({\dzg\over\HH}\right)=-V-{1+z\over H}\phi'-{1+z\over H}
{\partial V\over\partial r} \nonumber \\
&&-\dzg+{1+z\over H}{dH\over dz}~\dzg~,
\enar
where integration by part is performed in Eq.~(\ref{eq:dl}) and
we have ignored quantities at origin that can be absorbed to the observed
mean \cite{YOFIZA09,YOO10}.
The total derivative in Eq.~(\ref{eq:zdist}) is with respect to the
observed redshift~$z$ and it is related to the null geodesic as
\beeq
{d\over dz}={1\over H}{d\over dr}=-{1\over H}\left({\partial\over\partial\tau}
-{\partial\over\partial r}\right)~.
\label{eq:null}
\eneq
The derivative term also appeared as a partial derivative in
\cite{YOFIZA09,YOO10}, while we wrote it here as a total derivative to imply
Eq.~(\ref{eq:null}). However, in a sense
it is a partial derivative with respect to
the observed redshift with other observable quantities $(\theta,\phi)$
kept fixed.

The observed galaxy fluctuation in Eq.~(\ref{eq:fullGR}) is the sum of the
matter density~$\dm$, the gravitational potential~$\phi$, the line-of-sight
velocity~$V$, and other distortions
such as $\dzg$, $\drg$, and $\ddL$, and they are also a linear combination of
$\dm$, $\phi$, and $V$ with various prefactors and integrals
in Eqs.~(\ref{eq:los})$-$(\ref{eq:zdist}), which in turn can be expressed
in terms of the matter density~$\dm$. Using the Einstein equations, we have
\bear
\phi&=&{3H_0^2\over2}~{\OM\over ak^2}~\dm ~,\\
v&=&-{1\over k}~\dm'=-{\HH f\over k}~\dm~,\\
V&=&i\HH f~{\dm\over k}~\mu_k~,
\enar
where the logarithmic growth rate is $f=d\ln\dm/d\ln a$,
$d/d\tau=\HH d/d\ln a$, and
$\mu_k$ is the cosine angle between the line-of-sight direction
and the wavevector.

\begin{figure}[t]
\centerline{\psfig{file=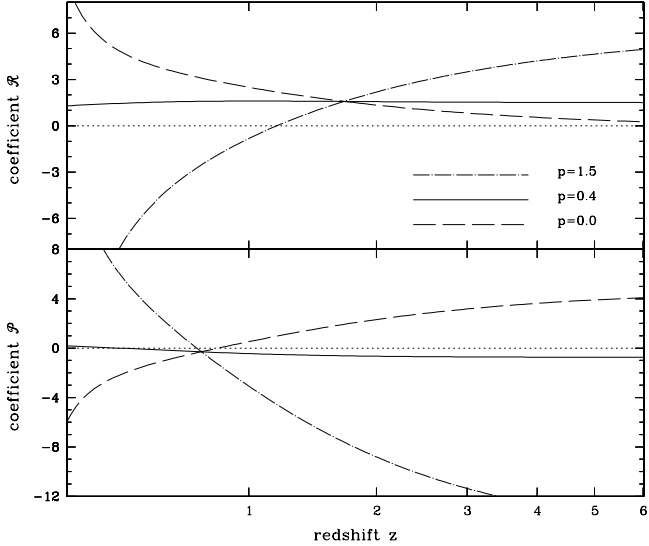, width=3.4in}}
\caption{Redshift dependence of two dimensionless parameters $\cB$ and $\cA$
in Eq.~(\ref{eq:PR}). Non-vanishing values of $\cB$ and $\cA$ represent the
general relativistic effects in galaxy clustering, each of which describes
the contributions of the gravitational potential and the velocity to the
observed galaxy fluctuation field.
Three different curves represent galaxy samples in a volume-limited 
survey ($p$ is constant) with three different limits $L_t$ in luminosity
threshold: a sample with low threshold $L_t\ll L_\star$ ($p=0$; dashed), 
a sample with no magnification bias $L_t\simeq L_\star$ $(p=0.4$; solid), 
a sample at high luminosity tail $L_t\gg L_\star$ ($p=1.5$; dot-dashed). 
The evolution bias factor $\ee=3$ is fixed in all cases,
representing homogeneous galaxy samples (constant comoving number density)
often constructed in large-scale galaxy surveys.}
\label{fig:PR}
\end{figure}

Before we compute the observed galaxy power spectrum, we make
a further simplification by ignoring the
projected quantities such as the gravitational lensing and the integrated 
Sachs-Wolfe contributions in Eqs.~(\ref{eq:dz})$-$(\ref{eq:dl}),
which are important only 
for the pure transverse modes ($k^\parallel=0$) \cite{HUGALO08,YOO10}.
With this simplification and by using Eqs.~(\ref{eq:los})$-$(\ref{eq:zdist}),
the observed galaxy fluctuation in Eq.~(\ref{eq:fullGR}) can be written 
in Fourier space as
\beeq
\dgr=\int{d^3\kvec\over(2\pi)^3}~e^{i\kvec\cdot\xvec}
\left[\dnewt+{\cA~\dm\over(k/\HH)^2}-i\mu_k{\cB~\dm\over k/\HH}\right]~,
\label{eq:full}
\eneq
where the two dimensionless coefficients $\cB$ and~$\cA$ 
are defined as
\bear
\label{eq:PR}
\cA&=& \ee f-{3\over2}\OM(z)\bigg[\ee+f-{1+\zz\over H}{dH\over dz}
\nonumber\\
&&+(5p-2)\left(2-{1\over\HH\rs}\right)\bigg] ~,\\
\cB&=&f\left[\ee-{1+\zz\over H}{dH\over dz}+(5p-2)\left(1-{1\over\HH\rs}
\right)\right] ~,\nonumber
\enar
and the standard Kaiser formula of the observed galaxy fluctuation is 
\beeq
\dnewt=b~\dm-\mu_k^2~{kv\over\HH}=(b+f\mu_k^2)~\dm~,
\label{eq:kaiser}
\eneq
which can be contrasted with $\dgr$ in Eq.~(\ref{eq:fullGR}).

Apparent from their spatial dependence in Eq.~(\ref{eq:full}),
the coefficients $\cB$~and~$\cA$
originate from the velocity and gravitational potential.
While Eq.~(\ref{eq:fullGR}) can be derived with the minimal
assumption that the spacetime is described by a perturbed FLRW metric and
photons follow geodesics, the coefficient~$\cA$ in Eq.~(\ref{eq:PR})
is obtained by applying the Einstein equations 
[Eqs.~(\ref{eq:poisson})$-$(\ref{eq:pot})]. Given the observed~$\ee$ and~$p$,
the value and the functional form
of~$\cA$ is, therefore, heavily dependent upon the general theory of 
relativity.

While $\cA$ is purely relativistic, some
contributions to~$\cB$ may be considered non-relativistic, since they 
could be written down in Newtonian dynamics,
simply as a coupling of velocity from the Doppler effect with
the time evolution of the galaxy number density.
However, different gravity models will yield different values of~$\cB$ via
the logarithmic growth rate~$f$ (see, e.g., \cite{REMAET10}, and
further discussion on this issue is 
presented in Sec.~\ref{sec:newt}).
Therefore, measuring~$\cB$ and~$\cA$ is equivalent to a direct 
measurement of the relativistic contributions, and
we collectively refer to the coefficients~$\cB$ and~$\cA$
as the general relativistic effects in the galaxy power spectrum. 

Figure~\ref{fig:PR} illustrates the redshift dependence of two coefficients
$\cB$ and $\cA$, in which the evolution bias factor is fixed $\ee=3$
(constant comoving number density).
The contributions to these two coefficients arise from the volume and the
source distortions that also involve the change in the observed redshift and
the observed flux. Since the volume distortion involves $2\drg/\rs$ in
Eq.~(\ref{eq:fullGR}), both coefficients
diverge at $z\rightarrow0$ ($r\rightarrow0$),
unless the radial distortion of the volume effect
is cancelled by the source effect ($p=0.4$), leaving only the distortion terms
from the observed redshift (as we show below this cancellation is generic). 
With $\ee=3$ and $p=0.4$, the coefficient~$\cA$
in Eq.~(\ref{eq:PR}) nearly vanishes at low redshift.

However, as we quantify in Sec.~\ref{sec:signi},
even if the divergent term does not cancel out, 
since the survey volume decreases faster, the diverging terms have negligible
impacts on the measurement significance.
Furthermore, Figure~\ref{fig:PR} appears
different from those obtained 
in \cite{JESCHI11}, since they adopted the halo model to relate the
evolution bias factor~$\ee$ to the galaxy bias factor~$b$. 
Large-scale galaxy surveys show that
these two parameters for galaxy samples are independent of each other,
and their relation is
different from the halo model prediction (e.g., \cite{COEIET08}).

Before we close this section, we discuss the fluctuation in the luminosity
distance.
Additional contribution $5p~\ddL$ to the observed galaxy fluctuation arises,
because galaxy samples are selected given an observational threshold in flux 
or equivalently a luminosity threshold~$L_t$ converted by using the 
observed redshift~$z$. For flux-limited samples, the luminosity threshold is 
changing as a function of redshift, as only brightest galaxies at high
redshift can have observed fluxes large enough to be above the threshold in
flux. Volume limited samples
are constructed by imposing a constant luminosity threshold~$L_t$, up to
a maximum redshift. Furthermore, the evolution bias factor~$\ee$ in this
case is defined with respect to the number density of galaxies with
$L>L_t$, in addition to other criteria such as color 
cuts.\footnote{The luminosity function naturally evolves
in time due to aging stars in galaxies, galaxy mergers, cosmic expansion,
and so on. However, the condition that
$\ee=3$ and $p$~is constant relies on the assumption that the shape of
the luminosity function at constant $L_t$ remains unchanged, while the
physical number density $\bnobs(>L_t)$ is diluted as due to cosmic expansion.}

Fluctuations in luminosity distance bring galaxies around the threshold $L_t$ 
into (or out of) the galaxy samples, and this effect provides the 
contribution $5p~\ddL$ in Eq.~(\ref{eq:fullGR}).
The luminosity function slope of source galaxy populations at the threshold
is defined as
\beeq
p={d\log\bnobs\over dM}=-0.4~{d\log\bnobs\over d\log L}=-0.4~\alpha_L~,
\label{eq:slope}
\eneq
where the absolute magnitude is related to the luminosity
as $M-M_\star=-2.5\log(L/L_\star)$ with pivot points $M_\star$ and
$L_\star$. 

The galaxy luminosity function is often described by the Schechter function
\cite{SCHEC76} with a power-law slope~$\alpha_s$ and an exponential cut-off as
\beeq
\phi(L)dL={\phi_\star\over L_\star}\left({L\over L_\star}\right)^{\alpha_s}
\exp\left[-{L\over L_\star}\right]~dL~,
\eneq
and the number density given a threshold in luminosity is then
\beeq
\bnobs(>L_t)=\phi_\star\Gamma(\alpha_s+1,L_t/L_\star)~,
\eneq
where $\Gamma(a,x)$ is the incomplete Gamma function. Therefore, 
the luminosity function slope at threshold is
\beeq
p=0.4~{(L_t/L_\star)^{\alpha_s+1}\exp\left[-{L_t\over L_\star}\right]\over 
\Gamma\left(\alpha_s+1,{L_t/L_\star}\right)}~,
\eneq
which can take values from zero to infinity, depending on the choice of~$L_t$.

A typical case of interest is a flux limited survey, which is dominated by 
$L_\star$-galaxies 
at the peak of the luminosity function. For these galaxy samples
$p=0.4$ (solid), which can be obtained with $L_t=L_\star$ 
and $\alpha_s=0$ (or their variants). 
These galaxy samples have no diverging terms in~$\cB$ and~$\cA$,
because the volume distortion is balanced by the fluctuation in the 
luminosity distance. 
We may also assume volume-limited galaxy samples with constant~$p$ and consider
two additional representative cases in Figure~\ref{fig:PR}.
First, galaxy samples at high luminosity tail ($p=1.5$; dot-dashed), 
here taken as
$L_t=3L_\star$ and for which the Schechter function slope is 
$\alpha_s\simeq-1.1$ (e.g., \cite{COEIET08}).
Last, we consider galaxy samples with sufficiently 
low threshold $L_t\ll L_\star$ 
($p=0$; dashed), which contain galaxies at low mass halos.

\section{Newtonian Correspondence and Redshift-Space Distortion}
\label{sec:newt}
In the cosmological context, full general relativistic equations reduce to
Newtonian equations on small scales, in which relativistic effects are
negligible. This statement is born out by observations that matter density
fluctuations~$\delta$
in various choices of gauge conditions become identical on small scales,
except insofar as unusual gauge conditions are imposed
such as the uniform density gauge, in which $\delta\equiv0$ on all scales.
However, as larger scale modes are considered, matter density fluctuations
become increasingly different from one another, and the transition scale is
largely set by horizon scales $k=\HH$.

Even on these large scales, however, ``Newtonian correspondence'' exists
in certain circumstances,
in the sense that the matter density~$\dm$, the velocity~$v$, and
the potential~$\phi$ fluctuations in Newtonian dynamics are identical to
those quantities in fully relativistic dynamics with certain choices of
gauge conditions \cite{HWNO99R,HWNO05}. In a flat universe with
presureless medium, the Newtonian matter density $\dm$ is identical to the
comoving gauge matter density~$\dv$, and the Newtonian velocity~$v$ and
potential~$\phi$ are identical to the conformal Newtonian gauge quantities
$\vx$ and $\px$. Since this correspondence holds on all scales, numerical
simulations properly capture large scale modes \cite{HWNO06,CHZA11},
apart from other technical issues such as finite box size.

In redshift-space distortion literature \cite{KAISE87,HAMIL98},
a Newtonian calculation has been performed for obtaining
the galaxy fluctuation field  $\delta_z$ in redshift-space that goes beyond 
the Kaiser formula in Eq.~(\ref{eq:kaiser}).
A different notational convention is often adopted in the redshift-space
distortion literature. Especially, galaxy number densities $n_g$
are expressed in comoving space.\footnote{Conversion of physical quantities
to quantities in comoving space requires a redshift parameter. In observation,
the observed redshift can be used without gauge ambiguity, but in general
a redshift parameter is defined in conjunction with the expansion factor
in a homogeneous universe. Therefore, it is a function of coordinate time 
and hence is gauge-dependent.
Only in this section, did we use $n_g$ to refer to
the galaxy number density in comoving space, as there is no gauge ambiguity
in the Newtonian limit.}
The selection function~$\alpha$ is defined with respect
to  the comoving number density $\bar n_g$  of galaxies as
\beeq
\alpha\equiv{d\ln r^2\bar n_g\over d\ln r}
=2+{rH\over1+z}\left(\ee-3\right)~,
\eneq
and the line-of-sight velocity $\VV$ is
\beeq
\VV\equiv{1+z\over H}~V\simeq{1+z\over H}~\dzg~,
\eneq
where
the last equality holds if we ignore potential contributions to $\dzg$
in Eq.~(\ref{eq:dz}). As a special case, the constant comoving number 
density~$\ee=3$ corresponds to $\alpha=2$, and both of them are constant.

The redshift-space distance~$s$ at the observed redshift~$z$ 
is then related to the ``real'' distance~$r$ as
\beeq
s\equiv\int_0^z{dz'\over H}=r+{1+z\over H}~\dz\simeq r+\VV~,
\eneq
where the first equality is exact to the linear order in perturbation,
while the
second equality neglects potential contributions
to~$\dz$.\footnote{Since the real distance~$r$ takes a gauge-dependent
redshift parameter
$\bar z$ as an argument, we left the gauge choice of $\dz$
unspecified, $1+z=(1+\bar z)(1+\dz)$.} From the conservation 
of total number of observed galaxies
$n_z(s)d^3s=n_r(r)d^3r$, one derives the relation between the redshift-space
and the real-space fluctuations as
\beeq
\label{eq:conr}
1+\delta_z={\bar n_g(r)\over \bar n_g(s)}\left|{d^3s\over d^3r}\right|
={r^2\bar n_g(r)\over s^2\bar n_g(s)}\left(1+{\partial\VV\over
\partial r}\right)^{-1}(1+b~\dm)~,
\eneq
and to the linear order in perturbations
the galaxy fluctuation in redshift-space is then
\bear
\label{eq:zfor}
\delta_z&=&b~\dm-\left({\partial\over\partial r}+{\alpha\over r}\right)\VV \\
&=&b~\dm-{1+z\over H}{\partial V\over \partial r}-e~V+2~V-{2V\over\HH r}
+{1+z\over H}{dH\over dz}~V~.\nonumber
\enar
This equation is known as the complete formula for the observed
galaxy fluctuation in the redshift-space distortion literature,
while only the first two terms constituting the Kaiser 
formula \cite{KAISE87} are often used. The additional terms 
come from the Jacobian of the transformation from real space to redshift
space and also from the galaxy number density evaluated at the observed
redshift.

Even with the knowledge of the Newtonian correspondence,
there are fewer terms in Eq.~(\ref{eq:zfor})
compared to the full relativistic formula $\dgr$ in
Eq.~(\ref{eq:fullGR}), and those terms account for missing physics
in the derivation. Apparently ignored are the fluctuation~$\ddL$
in the luminosity distance (effectively $p=0$) and 
the lensing contribution that
gives rise to distortions between the observed and the source angular
positions. 
Note that this by itself is an important effect: in a generic flux limited 
survey with $p=0.4$ the potentially divergent term ${\alpha\over r}\VV$ is 
exactly 
cancelled by the luminosity fluctuation effect, which is caused by the fact 
that 
a galaxy moved to a redshift closer to the observer will have its flux 
slightly smaller
than what the redshift distance predicts, so it may not enter into a flux 
threshold, which 
compensates for the volume effect in the $p=0.4$ case. 

However, once we account for luminosity threshold effects, the Newtonian
calculation can fully reproduce the velocity terms in Eq.~(\ref{eq:fullGR}).
While the velocity terms receive a relativistic contribution \cite{MCDON09} 
from the gradient of the potential in Eq.~(\ref{eq:zdist}), it is cancelled
by the time derivative of the velocity via the conservation (Euler)
equation in Eq.~(\ref{eq:cons}). Therefore, the functional form of~$V$ 
is generic in Eq.~(\ref{eq:fullGR}), same in all gravity theories
(the conservation equations should hold in other theories of modified gravity,
as it simply
states that the energy-momentum is locally conserved \cite{BERTS06}).
However, as we scale the velocity terms with the matter density~$\dm$ by
using the Poisson equation (and also the conservation equation), the value
of~$\cB$ itself will be different in other gravity theories than general
relativity, and its measurements can test general relativity,
although this kind of tests can be performed by using
the redshift-space distortion term in $\dnewt$.

With respect to the relativistic contributions,
adding the lensing contribution to the conservation relation
in Eq.~(\ref{eq:zfor}) is still not enough to recover the relativistic formula
in Eq.~(\ref{eq:fullGR}). While the lensing contribution accounts for
angular distortions, there exist a radial distortion in the source position,
the Sachs-Wolfe effect \cite{SAWO67} in the observed redshift, and
finally the difference in the observer and the galaxy rest frames.
Equation~(\ref{eq:zfor}) can be obtained from the relativistic formula
in Eq.~(\ref{eq:fullGR}) by ignoring potential contributions, assuming
$p=0$, and identifying the matter density and the line-of-sight velocity 
with those in the comoving and the conformal Newtonian gauges, respectively. 
However, the validity of the Newtonian description on large scales can only 
be judged retroactively, after the relativistic description is derived.
We emphasize again that 
a fully relativistic treatment is required for estimating $\cA$. 

\section{Multi-Tracer Shot Noise Cancelling Technique}
\label{sec:multi}
We consider multiple galaxy samples
with different bias factors for measuring
the general relativistic effects
in the galaxy power spectrum. Using a vector notation, the observed galaxy
fluctuation fields can be written as
\bear
\label{eq:ebias}
\bdobs&=&\left[\bb+f\mu_k^2\bI+{\ccA\boldsymbol{\cA}\over (k/\HH)^2}
-i\mu_k{\ccB\boldsymbol{\cB}\over k/\HH}\right]\dv+\bnn \nonumber\\
&\equiv&\BB(k,\mu_k)~\dv+\bnn~.
\enar
where $\bb$, $\bI$, $\bnn$ are the linear bias, the multi-dimensional 
identity, and the residual-noise field vectors. 
By definition the noise-field is independent of the
matter fluctuation $\AVE{\bnn\dv}=0$, and the square bracket in 
Eq.~(\ref{eq:ebias}) defines the effective bias vector~$\BB$.
We will adopt a plane parallel approximation for the power spectrum 
analysis, meaning there is only one angle $\mu_k$ between the Fourier mode 
and the line-of-sight direction
we need to consider. The corrections to this approximation 
are expected to be small \cite{SAPERA12}. 
More generally, the effects considered here are different from the 
plane parallel approximation and can be considered separately. 

The coefficients $\cA$ 
and~$\cB$ in Eq.~(\ref{eq:PR}) are also generalized to the multi-tracer case as
\bear
\boldsymbol{\cA}&=&
\bdv{\ee} f-{3\over2}\OM(z)\bigg[\bdv{\ee}+
\left(f-{1+\zz\over H}{dH\over dz}\right)\bI \nonumber \\
&&+(5\bdv{p}-2\bI)\left(2-{1\over\HH\rs}\right)\bigg]~,\\
\boldsymbol{\cB}&=&
f\left[\bdv{\ee}-{1+\zz\over H}{dH\over dz}\bI+(5\bdv{p}-2\bI)
\left(1-{1\over\HH\rs}\right)\right]~. \nonumber
\enar
We introduced two new parameters~$\ccB$ and~$\ccA$ to generalize
the measurement significance of the coefficients~$\cB$ and~$\cA$ to the case
of multiple galaxy samples --- they are $\ccB=\ccA=1$ in general 
relativity, and measurements of these two parameters amount to the
measurement significance of the two vectors~$\boldsymbol{\cB}$ 
and~$\boldsymbol{\cA}$.

In order to assess our ability to measure the general relativistic effects
in the galaxy power spectrum, we employ the Fisher information matrix,
and the likelihood of the measurements is 
\beeq
\mathcal{L}={1\over(2\pi)^{N/2}~(\up{det}\CC)^{1/2}} ~
\exp\left[-{1\over2}~\bdobsT\CCI\bdobs\right]~,
\eneq
where the covariance matrix is $\CC=\AVE{\bdobs\bdobsT}=\BB\BT\Pv+\EE$,
the shot noise matrix is $\EE=\AVE{\bnn\bnn^\up{T}}$, and 
the matter power spectrum in the comoving gauge is $\Pv(k)$.
Since the observed galaxy fluctuation fields are constructed to have a 
vanishing mean $\AVE{\bdobs}=0$, the Fisher information matrix is 
\beeq
F_{ij}=\left\langle-{\partial^2\ln\mathcal{L}\over\partial \theta_i
\partial \theta_j}
\right\rangle={1\over2}~\up{Tr}\left[\CCI\CC_i\CCI\CC_j\right]~,
\eneq
with two measurement significance parameters $\theta_i=\ccA$, $\ccB$
and two nuisance parameter vectors $\theta_i=\bee$, $\bdv{p}$.
The covariance matrix with subscript is $\CC_i=\partial\CC/\partial\theta_i$.

The inverse covariance matrix of the multi-tracer field 
and the derivative of the covariance matrix are
\bear
\CCI&=&\EE^{-1}-{\EE^{-1}\BB\BT\EE^{-1}\Pv\over1+\mA}~, \\
\CC_i&=&{\partial\CC\over\partial\theta_i}=\left(\BB_i\BT+\BB\BT_i\right)
\Pv~, \nonumber
\enar
where $\mA=\BT\EEI\BB\Pv$~, $\BB_i=\partial\BB/\partial\theta_i$~, and 
we ignored the derivative of the shot noise matrix. 
With the inverse covariance matrix, we have
\bear
&&\CCI\CC_i\CCI\CC_j=\Pv^2\Big(\CCI\BB\BT_i\CCI\BB\BT_j
+\CCI\BB\BT_i\CCI\BB_j\BT \nonumber \\
&&\hspace{10pt}
+\CCI\BB_i\BT\CCI\BB\BT_j+\CCI\BB_i\BT\CCI\BB_j\BT\Big)  ~, 
\enar
and the Fisher information matrix is
\bear
\label{eq:intm}
F_{ij}&=&\left(\BT\CCI\BB\right)\up{Re}\left(\BT_i\CCI\BB_j\right)\Pv^2 \\
&&
+\up{Re}\left[\left(\BT\CCI\BB_i\right)\left(\BT\CCI\BB_j\right)\right]
\Pv^2~. \nonumber
\enar
To further simplify the equation, we define two more coefficients
\bear
\mB_i&=&\BT\EEI\BB_i\Pv~,\\
\mC_{ij}&=&\up{Re}\left(\BT_i\EEI\BB_j\right)\Pv={\BT_i\EEI\BB_j+\BT_j\EEI\BB_i
\over2}~\Pv~, \nonumber
\enar
and the various terms in Eq.~(\ref{eq:intm}) are
\bear
\BT\CCI\BB\Pv&=&{\mA\over1+\mA}~,\\
\BT\CCI\BB_i\Pv&=&{\mB_i\over1+\mA}~,\nonumber \\
\up{Re}\left(\BT_i\CCI\BB_j\right)\Pv&=&
\mC_{ij}-{\mBS_i\mB_j+\mBS_j\mB_i\over2~(1+\mA)}~. \nonumber
\enar
The Fisher information matrix is therefore
\beeq
\label{eq:fisher}
F_{ij}={\mA\over1+\mA}~\mC_{ij}+
{\up{Re}\left(\mB_i\mB_j-\mA\mB_i\mBS_j\right)\over(1+\mA)^2}~.
\eneq
This formula is a straightforward extension of the Fisher matrix in
\cite{HASEDE11} with the effective bias vector~$\BB$ being a complex vector.
The imaginary part arises solely from the $\cB$-term in Eq.~(\ref{eq:ebias})
and its derivative.

\begin{figure}[t]
\centerline{\psfig{file=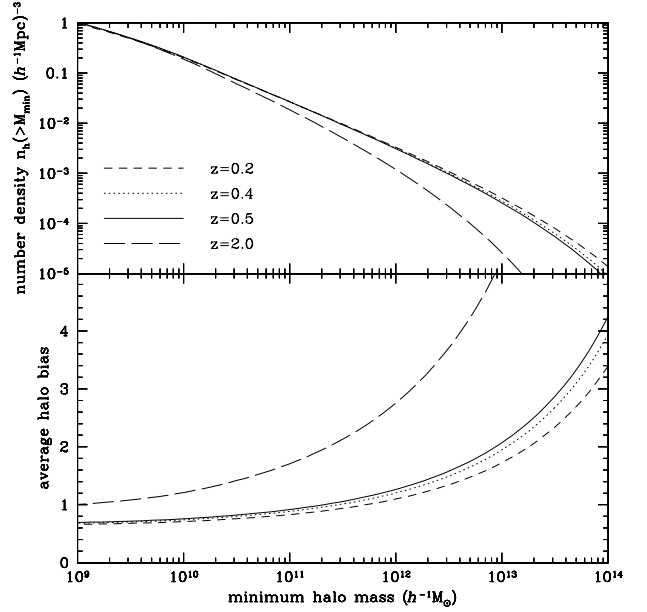, width=3.3in}}\vspace{-10pt}
\caption{Number density and average bias
of halos above the minimum mass at different 
redshift slices. Since the multi-tracer method utilizes all the halos of
mass above the minimum mass, a large number of halos are required to achieve
sufficiently low minimum mass.}
\label{fig:nn}
\end{figure}

For comparison, we also consider measurements by using a single galaxy sample.
The formalism presented in Sec.~\ref{sec:multi} is valid for a single tracer,
in which vector quantities reduce to scalar quantities.
The Fisher information matrix for a single tracer is
\beeq
F_{ij}={2~\up{Re}(\mB_i)\up{Re}(\mB_j)\over(1+\mA)^2}~,
\eneq
where we used the relation
\beeq
\mA\mC_{ij}={\mB_i\mBS_j+\mBS_i\mB_j\over2}~,
\eneq
only valid for a single tracer.
Note that in the single tracer we are not sensitive to the correlation between 
real and imaginary part of the mode, which has the dominant contribution 
to the signal-to-noise ratio of~$\cB$
in multi-tracer method \cite{MCDON09}. This is only true in the 
plane parallel approximation. 

\begin{figure*}[t]
\centerline{\psfig{file=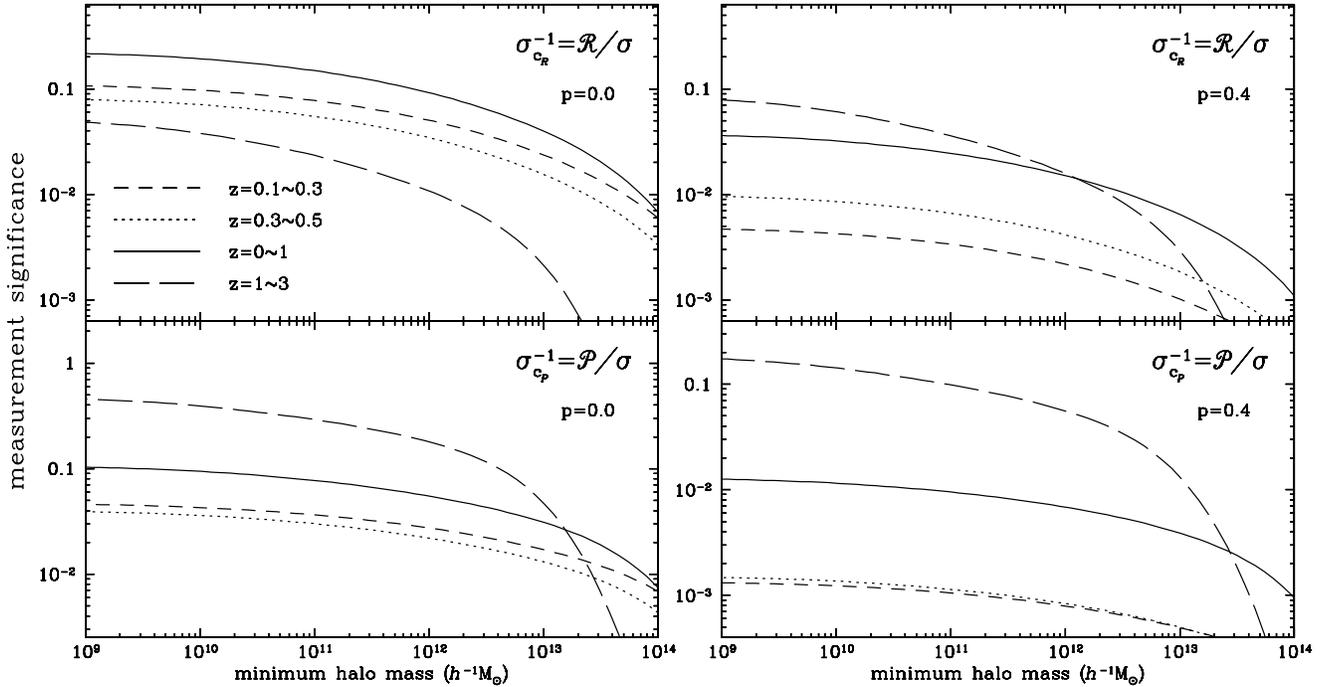, width=4.3in, angle=-90}}\vspace{-10pt}
\caption{Predicted measurement significance of general relativistic
effects $\cB$ (upper) and $\cA$ (bottom) in the galaxy power spectrum
obtained by using a single tracer. 
All halos of mass above minimum mass are lumped
together to construct a  single tracer. Four curves represent 
different survey redshift ranges with corresponding 
volume $V=2.5$, 7.9, 59, 410~$(\hgpc)^3$. 
For the volume-limited sample (constant $p$)
with constant comoving number density ($\ee=3$),
two galaxy samples are constructed to have $p=0$ (left) and $p=0.4$ (right).
No uncertainties in theoretical predictions are assumed.
With the traditional power spectrum analysis (single tracer), it is difficult
to measure the general relativistic effects at any meaningful significance.}
\label{fig:single}
\end{figure*}

\section{Measurement Significance} 
\label{sec:signi}
For definiteness, we consider full sky
surveys with three different redshift ranges and adopt a set of
cosmological parameters consistent with the WMAP7 results \cite{KOSMET11}.
Given the survey volume~$V$, 
the Fisher matrix is summed over the Fourier volume, 
where $\kmin=2\pi/V^{1/3}$ and  $\kmax=0.03\hmpci$ (we clarify the dependence
of the measurement significance on our choice of $\kmin$ and $\kmax$).
To model the Fisher matrix parameters $\mA$, $\mB_i$, $\mC_{ij}$, we adopt the
halo model description in \cite{HASEET10,HASEDE11}; 
It has been well tested against
a suite of $N$-body simulations with Gaussian and non-Gaussian 
initial conditions. 

We assume that the galaxy samples are constructed to have a constant comoving
number density ($\bee=3\bI$) in a volume-limited survey (constant 
$\bpp=p\bI$). While uncertainties in $\bee$ and $\bpp$ can propagate 
to the measurement uncertainties in $\cA$ and $\cB$, we focus on how well
future surveys can measure $\cA$ and $\cB$, assuming
there are no uncertainties in theoretical predictions of $\cA$ and $\cB$.
Given the current measurement uncertainties in $\bee$ and $\bpp$ 
\cite{COEIET08}, the uncertainties in theoretical predictions are very small
and will be smaller in future surveys.
Figure~\ref{fig:nn} shows the number density and average
bias of halos of mass
above the minimum mass $\Mmin$ at different redshift slices.
As we are interested in applying the multi-tracer method
with sufficiently low minimum mass to enhance the measurement significance
of the general relativistic effects, we consider two cases for the 
luminosity function slope at the threshold: $p=0$ (sufficiently low 
threshold $L_t\ll L_\star$) and $p=0.4$ ($L_t\simeq L_\star$).

\subsection{Single tracers}

First, we consider the prospect of measuring the general relativistic
effects by using a single tracer. Figure~\ref{fig:single} shows the
predicted measurement significance of~$\cB$ and~$\cA$ for various survey
redshift ranges. For all galaxy samples with different minimum mass
(or different number density in Figure~\ref{fig:nn}), the predicted 
measurement significance is very weak, indicating that 
substantial difficulty is present in measuring the general 
relativistic effects in the galaxy power spectrum for surveys at $z<3$. 
This difficulty is simply due to the small number of large-scale
modes that are sensitive to the general relativistic effects.
Furthermore, the weak measurement significance of both~$\cB$ and~$\cA$
means that compared to the standard Newtonian contribution~$\dnewt$,
the general relativistic effects or additional terms in~$\delta_z$ 
[Eq.~(\ref{eq:zfor})] used in the redshift-space distortion literature
are all negligible in the standard power spectrum analysis (single tracer
method). 

This result confirms the prediction in \cite{YOO10}
and extends the predictions to galaxy samples with different number 
density and bias. This conclusion is in apparent contradiction with 
\cite{RASAPE10}, where correlation functions are shown with smaller 
errors than the size of the effects. However, these errors are obtained from 
simulations without taking into account the actual number of modes in a 
realistic survey. Once this is taken into account there is no contradiction 
\cite{SAPERA12}. The reason the wide-angle correction remains small
is that there exists no velocity-density correlation ($\sim\cB$)
due to symmetry of pairs in Eq.~(\ref{eq:full}) and the effect shows up only 
as a correction to the dominant contributions that are already accounted 
in the plane-parallel limit.

\begin{figure*}[t]
\centerline{\psfig{file=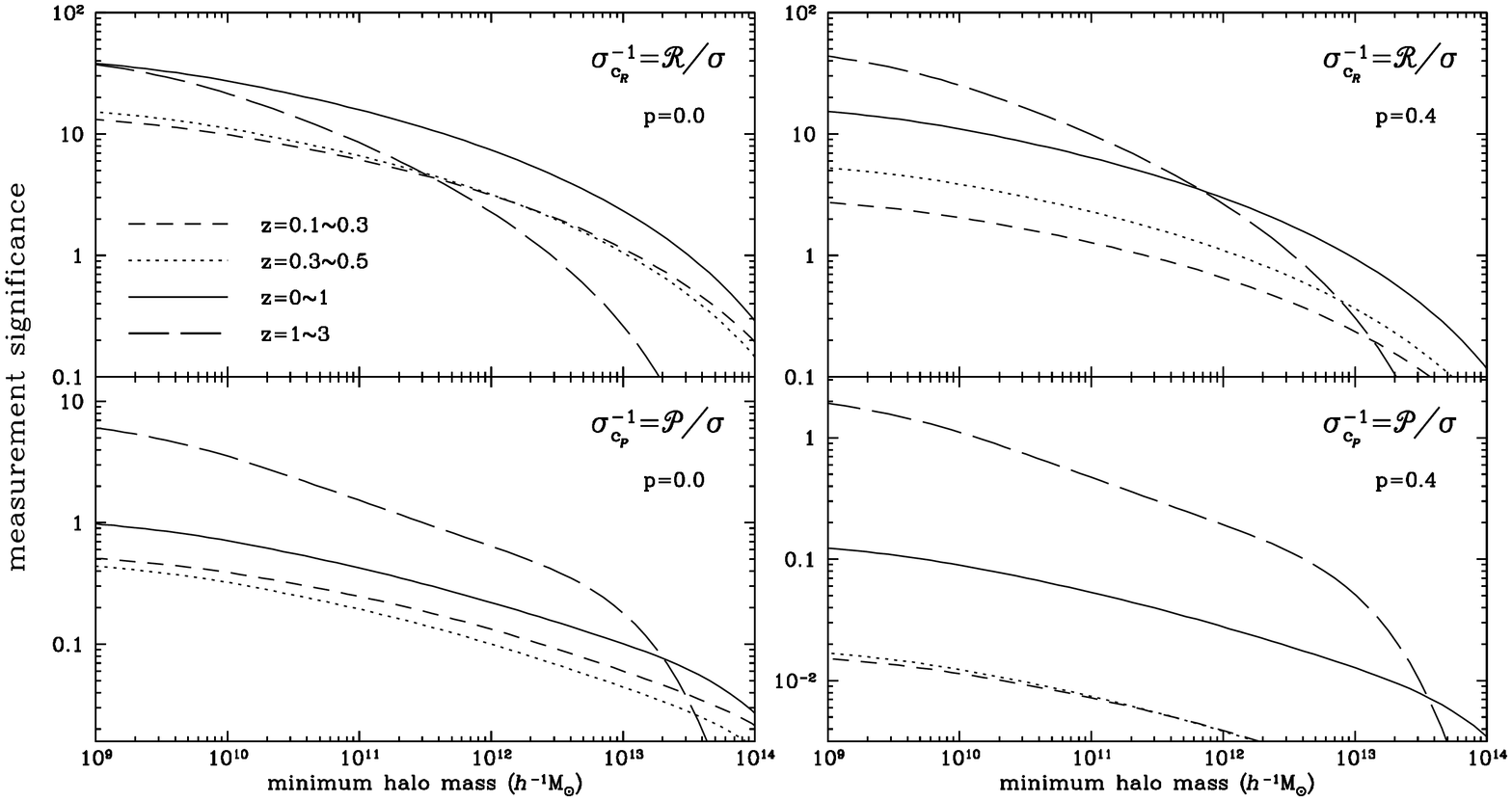, width=4.3in, angle=-90}}\vspace{-10pt}
\caption{Predicted measurement significance of general relativistic
effects $\cB$ (upper) and $\cA$ (bottom) in the galaxy power spectrum
derived by using the multi-tracer method. All halos of mass above 
minimum mass are utilized to take advantage of the multi-tracer
method \cite{SELJA09,HASEDE11}. Various curves are in the same format as
in Figure~\ref{fig:single}. Compared to Figure~\ref{fig:single}, 
the measurement significance is substantially enhanced by using
the multi-tracer method.}
\label{fig:sig}
\end{figure*}

The difference in the predicted measurement significance between the left 
($p=0.0$) and the right ($p=0.4$) panels arise from the redshift dependence
of the~$\cB$ and~$\cA$ values, as illustrated in Figure~\ref{fig:PR}.
The signals are computed at the mean redshift of each survey,
and in the left panel ($p=0.0$), the absolute values of
$\cB$ and~$\cA$ slowly decrease with redshift, while they remain nearly
constant in the right panel ($p=0.4$). Furthermore, even with higher 
suppression power of $(k/\HH)$, it is generally easier to measure~$\cA$ 
than~$\cB$ in the single tracer method for the case with $p=0.0$ (left panel).
Arising from the imaginary part in Eq.~(\ref{eq:ebias}),
the $\cB$-term in the galaxy power spectrum is negligible,
compared to the real part that includes the standard Newtonian term and
the~$\cA$-term, while the sensitivity to the $\cA$-term comes from the
cross-correlation of the standard Newtonian term and the $\cA$-term.
In the left panel, the lack of sensitivity to the $\cA$-term is due to
the vanishing signal of~$\cA$.

While the shot noise of massive halos
is an obstacle for measuring the general relativistic effects, it is the
cosmic variance on large scales that fundamentally 
limits the measurement significance. Therefore, the measurement significance
slowly increases as the minimum halo mass is lowered, but
it quickly saturates given values of~$\cB$ and~$\cA$. 
For measuring the general relativistic effects in the galaxy power spectrum,
there is no further gain in 
constructing galaxy samples with large number density at a fixed survey 
volume.

\subsection{Multiple tracers}

This conclusion is, however, contingent upon the assumption that the power
spectrum analysis is performed by using a single tracer, and the multi-tracer
method can change the prospect of measuring the general relativistic effects
in a dramatic way.
Figure~\ref{fig:sig} shows the predicted measurement significance
of the general relativistic effects by taking full advantage of the
multi-tracer method. Compared to the single tracer case in 
Figure~\ref{fig:single}, there exist two key differences in the measurement
significance derived by using the multi-tracer method. First, the measurement
significance is greatly enhanced in Figure~\ref{fig:sig} by eliminating 
the cosmic variance, which sets the fundamental limit in the single tracer
method. Second, a substantially larger measurement significance of 
the $\cB$-term is 
obtained than that of the $\cA$-term by cross-correlating multiple galaxy
samples and thereby isolating the imaginary term \cite{MCDON09}.
The method of measuring the imaginary part in 
the galaxy power spectrum of two tracers \cite{MCDON09} is fully
implemented in our complex covariance matrix as off-diagonal terms
and extended to the number of tracers larger than two. 
The result in \cite{MCDON09} 
would correspond to  $\cB/\sigma\simeq1.8$ at $z\leq1$. 

In our most optimistic scenario, 
if we can utilize all halos of $M\geq10^{12}\hmsun$, 
the velocity term~$\cB$  (solid) of the galaxy samples with $p=0$ (left panel) 
can be measured  at more than 10-$\sigma$ confidence level
at $z\leq1$, while it is still difficult to detect
the gravitational potential term~$\cA$ (solid).
A significant detection of~$\cB$ can be made, even in surveys at low redshift 
(dotted and short-dashed), if halos of $M<10^{12}\hmsun$ can be used.
At higher redshift $z\geq1$, though the increase in the survey volume 
is partially cancelled by the lower 
abundance of halos at a fixed mass, a substantial improvement (dashed)
for~$\cA$ can be achieved by going beyond $z=1$, as the signal~$\cA$ increases
with redshift ($p=0$).

However, the scenario above is not very realistic because of 
the $p=0$ assumption. 
In the right panels, we consider the galaxy samples with $p=0.4$, of which
the~$\cB$ and~$\cA$ values are nearly constant at all redshifts. The constant
signals result in higher measurement significance for surveys with larger
volume at higher redshift. Compared to the case with $p=0.0$, 
the measurement significance of~$\cB$ is smaller due to the smaller 
value of~$\cB$ at $z<1$, while that of~$\cA$ is highly suppressed
due to the vanishing value of~$\cA$.
By using halos of mass slightly lower than $10^{12}\hmsun$,
a survey like the BOSS that covers redshift range $z=0.3\sim0.5$ with a
quarter of the sky can achieve a 1-$\sigma$ detection of~$\cB$,
demonstrating the feasibility of the multi-tracer analysis
in future surveys.

A few caveats are in order. First, we used the galaxy
power spectrum in a flat-sky and counted the number of modes in computing
the measurement significance. Calculations of wide-angle correlation functions
\cite{SAPERA12} 
shows that the effects of geometry are negligible, and
a similar calculation can be performed for power spectrum measurements
(in preparation). Furthermore,
our calculation is rather insensitive to the minimum wavenumber $\kmin$.
At sufficiently low $\Mmin$, the multi-tracer method approaches the optimal
case with dark matter density field, where the Fisher information can be
approximated as $\BB'\EE^{-1}\BB'\Pv$. In terms of spatial dependence alone
in Eq.~(\ref{eq:ebias}),
$F_{\cA}\propto\int~ dk~ k^2(k^n/k^4)\propto\ln(\kmax/\kmin)$ and
$F_{\cB}\propto\int dk~ k^2(k^n/k^2)\propto (\kmax^2 - \kmin^2)$ 
with the spectral index $n\simeq1$. Therefore, the dependence of the
measurement significance on $\kmin$ is logarithmic for~$\cA$
and negligible for~$\cB$, as $\kmax\gg\kmin$. In contrast, a significance
enhancement in $F_\cB$ can be achieved by increasing
the maximum wavenumber $\kmax$, although the gain is marginal for~$F_\cA$.
Finally, while any degeneracy with cosmological
parameters in measuring the relativistic effects is largely eliminated due to
the cancellation of 
the underlying matter distribution \cite{SELJA09,HASEDE11},
a proper quantification of the measurement significance requires
considerations of uncertainties in theoretical predictions of $\cA$ and
$\cB$.

\section{Primordial non-Gaussianity} 
\label{sec:fnl}
We extend our formalism 
to the primordial non-Gaussian signature in galaxy bias \cite{DADOET08} and
introduce additional parameter~$\fnl$.
Here we only consider the simplest local form of primordial non-Gaussianity
to demonstrate how it can be implemented in the general relativistic 
description, and ignore scale-independent and scale-dependent corrections
(see, e.g., \cite{SLHIET08,DESE10,DEJESC11a}).

The primordial non-Gaussian signature in galaxy bias can be readily 
implemented in our full general relativistic description with the Gaussian
bias factor in Eq.~(\ref{eq:fullGR}) replaced by
\beeq
b\rightarrow b+3\fnl(b-1)\delta_c~{\OM(z)\HH^2\over T_\varphi(k,z)~k^2}~,
\label{eq:fnl}
\eneq
where $\delta_c$ is the linear overdensity for spherical collapse and
$T_\varphi$ is the transfer function of the
curvature perturbation (see also \cite{BRCRET11,BASEET11,JESCHI11}).
Equivalently, the primordial non-Gaussianity can be considered as additional
relativistic contribution by replacing $\cA$ in Eq.~(\ref{eq:PR}) with
\beeq
\cA_{\fnl}=\cA+3\fnl(b-1)\delta_c~{\OM(z)\over T_\varphi(k,z)}~.
\eneq

\begin{figure}[t]
\centerline{\psfig{file=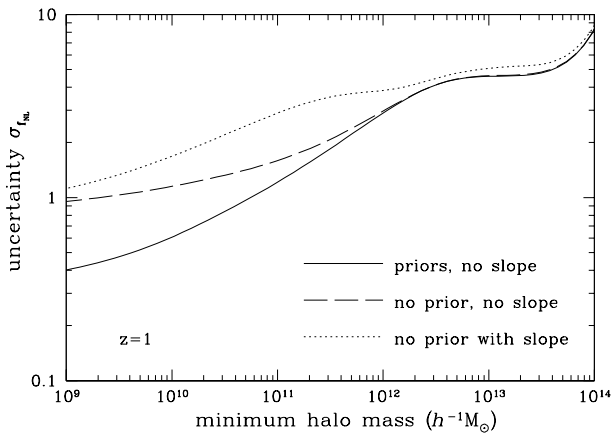, width=3.3in}}\vspace{-10pt}
\caption{Predicted constraints on the primordial non-Gaussianity $\fnl$
from galaxy power spectrum measurements. To facilitate the comparison,
the constraints on $\fnl$ are obtained by using the same survey specifications 
as in \cite{HASEDE11}: $V\simeq50~(\hgpc)^3$ centered at $z=1$. We assume
$\ee=3$ and $p=0.4$, and various
curves show $\sigma_{\fnl}$ with different priors on $\bee$ and $\bpp$
($\sigma_\ee=0.1$, $\sigma_p=0.05$ \cite{COEIET08} for the
solid curve).}
\label{fig:fnl}
\end{figure}

Figure~\ref{fig:fnl} shows the predicted constraints on the primordial
non-Gaussianity derived by accounting for the full general relativistic
effects.
In obtaining the constraint $\sigma_{\fnl}$ on primordial non-Gaussianity,
we set $\ccB=\ccA=1$ and marginalize over~$\ee$ and~$p$
with priors $\sigma_\ee=0.1$
and $\sigma_p=0.05$ \cite{COEIET08}. We emphasize that $\ee$ and $p$
can be more accurately measured in observations, further reducing their 
uncertainties. With the
current uncertainties in~$\ee$ and~$p$, the constraint $\sigma_{\fnl}$
(solid in Fig.~\ref{fig:fnl}) is nearly identical
to the unmarginalized constraint. 
The dashed curve shows that even with no priors 
on~$\ee$ and~$p$, $\sigma_{\fnl}$ is not inflated
except in the regime with $\sigma_{\fnl}\lesssim2$, because $\cB$
and $\cA$ are affected simultaneously by~$\ee$ and~$p$
but only $\cA$ by $\fnl$. Furthermore, the unique dependence
of $\fnl$ on $b-1$ and $T_\varphi$ in Eq.~(\ref{eq:fnl}) provides the 
multi-tracer method with more leverage to separate it from the general
relativistic effect. It is the primordial gravitational potential
at initial epoch, not the evolved gravitational potential at the observed
redshift that the scale-dependent galaxy bias responds to, and the
difference is the transfer function~$T_\varphi(k,z)$. At low redshift,
the transfer function decays from unity on scales smaller than the horizon
around the dark energy domination epoch, 
and there exists a factor two difference in
$T_\varphi(k,z)$ at our adopted $\kmax$.

Finally, we allow~$\bee$ and~$\bpp$ to vary as a function of mass
with two logarithmic slope parameters $\alpha_\ee$ and $\alpha_p$:
\bear
\bee&=&\ee_0\bI+\alpha_\ee\ln(\boldsymbol{M}/M_0)~, \\
\bpp&=&p_0\bI+\alpha_p\ln(\boldsymbol{M}/M_0)~, \nonumber
\enar
with $\ee_0=3$, $p_0=0.4$,
$\alpha_\ee=\alpha_p=0$, and $M_0=10^{12}\msun$.
The effects of~$\alpha_e$ and~$\alpha_p$ (dotted in Fig.~\ref{fig:fnl})
are sufficiently different from that of~$\fnl$, and
$\sigma_{\fnl}$ asymptotically reaches the floor set by the uncertainties
in~$\ee$ and~$p$.
This demonstrates that the general relativistic effects in the
galaxy power spectrum are {\it not} degenerate with the primordial non-Gaussian
signature. However, if $\fnl$ were to be constrained below unity,
similar precision needs to be achieved in predicting~$\cB$ and~$\cA$.

This requirement can be relaxed by increasing the maximum wavenumber $\kmax$
to exploit the unique dependence of~$\fnl$ on $T_\varphi(k,z)$. 
With larger maximum wavenumber
$\kmax=0.1\hmpci$, the overall uncertainties on $\fnl$ are reduced by 
about 30\% for all three cases in Figure~\ref{fig:fnl}. Another way is to
construct galaxy samples by using independent mass estimates 
instead of observed flux,
as $p$ drops out in Eq.~(\ref{eq:full}), further separating $\ee$
from $\fnl$.

\section{Discussion}
\label{sec:discussion} 
In this paper we explore the contributions to the redshift survey beyond the 
Kaiser approximation in the context of recently developed general relativistic 
analysis. We compare the results of this formalism to the previous analyses  
\cite{RASAPE10} and 
show that these analyses ignore several terms such as luminosity distance 
fluctuation and overestimate the significance 
of the effect. In addition, correlation function analyses previously 
adopted are not optimal to assess the signal-to-noise ratio of these effects, 
in contrast to our power spectrum analysis. 
We find that these corrections beyond the Kaiser formula are not observable 
in a generic redshift survey using a single tracer, meaning that these effects 
do not have to be considered in a generic 
redshift survey. A caveat to this conclusion is that in this paper we have 
performed 
the analysis using the plane parallel approximation and do not consider 
large angle effects \cite{SZMALA98,SZAPU04}, 
but we expect that these effects are equally small \cite{SAPERA12}. 
Their 
detectability will be addressed in a separate publication. 

Using the multi-tracer shot noise cancelling method the detection significance 
is increased,
providing a unique opportunity to test these effects, and general 
relativity in general, 
on horizon scales. Still, for realistic cases the detection significance 
is at the few sigma confidence level (in some more optimistic cases the     
detectability rises to 10 sigma or more), 
so this test of general relativity is of interest only when 
the deviations from general relativity are significant on large scales. 
We have also shown how the primordial non-Gaussian effect in galaxy bias can
be implemented in the full general relativistic description, and 
we have argued that the ability to detect primordial non-Gaussianity is
little compromised by the presence of general relativistic effects.
 
Considering the fact that we perform a very large-scale analysis,
our method of measuring the general relativistic effects in galaxy 
clustering need not be restricted to spectroscopic surveys.
The use of photometric redshift
measurements may not affect our results if the photo-$z$ error is
sufficiently small, e.g. an error of $dz/(1+z)\sim 0.03$ 
corresponds to $k > 0.06 \hmpci$ at $z=1$, which is larger than 
$\kmax = 0.03 \hmpci$ we adopted here.
This allows one in principle to extend the observed halo mass ranges 
to lower masses using a photometric survey and to take full advantage of
the multi-tracer method.

While we treated multiple galaxy samples as halos in multiple mass bins
and our method requires sufficiently low $\Mmin$,
the SDSS already measures galaxies well below $M=10^{11}\hmsun$ 
(e.g., SDSS L1 sample), and there exists numerous methods to 
remove satellite galaxies and isolate central galaxies. Furthermore,
it is shown \cite{HASEDE11} that one needs a fairly large scatter in the 
mass-observable relation, $\sigma_{\ln M}=0.5$, to degrade 
the shot-noise suppression, and the scatter is less important at the low mass 
end. 
The reason for this insensitivity to the scatter is that
in terms of weighting, the optimal weighting method puts more weight on massive
   halos, $w(M) = M + M_0$, where $M_0$ is a constant and approximated
   as 3 times the minimum mass. 
 We note however that our prediction is based on the 
halo model description of the shot 
noise matrix, which is tested only for halos at $M\geq10^{12}\msun$, and our
prediction at $M\leq10^{12}\msun$ is an extrapolation. 

The bottom line of this paper is that the corrections beyond the Kaiser 
formula in redshift surveys 
are generally small and only detectable using 
very specialized techniques adopted in this paper. This is a good news 
for those analyzing 
generic redshift surveys since they do not have to consider them. 
Nevertheless,  the potential detectability of these terms 
gives rise to the prospect of testing general relativity in a 
regime previously untested. 
Thus despite the small detection significance it is 
worth 
exploring these tests further to see if they can be of use in separating 
general relativity from 
some of its alternatives. 

\acknowledgments
We acknowledge useful discussions with Daniel Eisenstein, Lucas Lombriser, 
and Pat McDonald.  J.Y. is supported by the SNF Ambizione Grant.
This work is supported by the Swiss National Foundation under contract
200021-116696/1 and WCU grant R32-10130.
M.~Z. is supported by the National Science Foundation under
PHY-0855425 and AST-0907969 and by the David and Lucile Packard foundation
and the John~D. and Catherine~T. MacArthur Foundation.

\vfill
\bibliography{detect.bbl}

\end{document}